\documentclass[12pt]{article}

\usepackage{amsmath}
\usepackage{amssymb}

\allowdisplaybreaks

\begin{document}

\baselineskip=18.8pt plus 0.2pt minus 0.1pt

\makeatletter

\@addtoreset{equation}{section}
\renewcommand{\theequation}{\thesection.\arabic{equation}}
\renewcommand{\thefootnote}{\fnsymbol{footnote}}
\newcommand{\beq}{\begin{equation}}
\newcommand{\eeq}{\end{equation}}
\newcommand{\bea}{\begin{eqnarray}}
\newcommand{\eea}{\end{eqnarray}}
\newcommand{\nn}{\nonumber\\}
\newcommand{\hs}[1]{\hspace{#1}}
\newcommand{\vs}[1]{\vspace{#1}}
\newcommand{\Half}{\frac{1}{2}}
\newcommand{\p}{\partial}
\newcommand{\ol}{\overline}
\newcommand{\wt}[1]{\widetilde{#1}}
\newcommand{\ap}{\alpha'}
\newcommand{\bra}[1]{\left\langle  #1 \right\vert }
\newcommand{\ket}[1]{\left\vert #1 \right\rangle }
\newcommand{\vev}[1]{\left\langle  #1 \right\rangle }
\newcommand{\ul}[1]{\underline{#1}}

\makeatother

\begin{titlepage}
\title{
\vspace{1cm}
On the Spectrum of $D=2$ Supersymmetric Yang-Mills Quantum Mechanics
}
\author{Yoji Michishita
\thanks{
{\tt michishita@edu.kagoshima-u.ac.jp}
}
\\[7pt]
{\it Department of Physics, Faculty of Education, Kagoshima University}\\
{\it Kagoshima, 890-0065, Japan}
}

\date{\normalsize March, 2011}
\maketitle
\thispagestyle{empty}

\begin{abstract}
\normalsize
We investigate the structure of the spectrum of states in $D=2$ SU($N$) supersymmetric Yang-Mills
matrix quantum mechanics, which is a simplified model of Matrix theory. We compute the thermal
partition function of this system and give evidence for the correctness of naively conjectured structure
of the spectrum. It also suggests that Claudson-Halpern-Samuel solution is the unique eigenfunction 
of simultaneously diagonalizable hermitian operators, and we show that it is true in $N=3$ and $N=4$ cases.
\end{abstract}
\end{titlepage}

\clearpage
\section{Introduction}

Matrix theory is expected to be a correct description of M-theory, and is given as 
an U($N$) matrix quantum mechanics. The structure of the spectrum of states of this theory is
complicated, and we know little about it except that the spectrum is continuous with some discrete poles.

Instead, in this paper we investigate $D=2$ SU($N$) supersymmetric Yang-Mills matrix quantum mechanics,
whose action is given by the dimensional
reduction of $D=2=1+1$ super Yang-Mills theory to $0+1$ dimension. As a simplified model of Matrix theory,
which is from $D=10$ super Yang-Mills theory, we expect that it gives some hint to the spectrum of
Matrix theory. This quantum mechanics is almost free, but Gauss law associated with the gauge fixing procedure
makes it nontrivial. 

The action of $D=2$ matrix quantum mechanics is given by
\bea
S & = & \frac{1}{g^2}\int dt\;\text{tr}\Big[
\frac{1}{2}DXDX+ig^2\theta^\dagger D\theta-g^4\theta^\dagger[X,\theta]
\Big],
\eea
where $g$ is the coupling constant, $DX=\p_tX-i[A_0,X]$, and $X, A_0, \theta$ are matrices 
in {\it su}($N$) Lie algebra. $\theta$ is also a complex Grassmann odd operator.
This action is invariant under the gauge transformation, and the following supersymmetry transformation:
\beq
\delta X=i(\epsilon^\dagger\theta+\epsilon\theta^\dagger),\quad
\delta\theta=-\Pi\epsilon,\quad
\delta A_0=g^2\delta X.
\eeq
Quantization of this system gives the following commutation relations:
\beq
[X^a,\Pi^b]=i\delta_{ab},\quad
\{\theta^a,(\theta^b)^\dagger\}=\delta_{ab},
\eeq
where $\Pi=\frac{1}{g^2}DX$ and $a$ is an index of SU($N$) adjoint representation.
Physical states must be gauge invariant: $G^a\ket{\text{phys}}=0$, where $G$ is the generator of SU($N$) :
\beq
G= i[X,\Pi]+\{\theta^\dagger,\theta\}.
\eeq
Hamiltonian $H$ and supercharge $Q$ are
\bea
H & = & g^2\text{tr}\;\Big[\frac{1}{2}\Pi^2\Big] + (\text{terms proportional to 1st class constraints}), \\
Q & = & -\text{tr}\;[\Pi\theta].
\eea
On physical states $H$ acts as free Hamiltonian $h=g^2\text{tr}\;[\frac{1}{2}\Pi^2]$.
However, this system is not trivial, because of the physical condition $G\ket{\text{phys}}=0$.
Commutation relations of $H$ and $Q$ are usual ones, up to operators which vanish on physical states:
\bea
\{Q^\dagger,Q\} & = & \frac{2}{g^2}h+(\text{terms proportional to 1st class constraints}),
\\
\{Q,H\} & = & (\text{terms proportional to 1st class constraints}),
\\
\{Q^\dagger,H\} & = & (\text{terms proportional to 1st class constraints}).
\eea

The fermion part of the states is constructed by acting $(\theta^a)^\dagger$ on a vacuum
$\ket{0_F}$ satisfying $\theta^a\ket{0_F}=0$. Then the total states take the following form:
\beq
(\theta^{a_1})^\dagger\dots(\theta^{a_n})^\dagger\ket{0_F}
\int\prod_adX^a\left[\prod_a\ket{X^a}\right]\psi(X).
\eeq

A gauge invariant solution to Schr\"odinger equation in the sector of zero fermion number
has been found in \cite{ch85,s07}: When $X$ is diagonalized by the gauge transformation:
\beq
X=U
\begin{pmatrix}
x_1 & 		& & \\
	& x_2 	& & \\
	&		& \ddots & \\
	&		&		& x_N
\end{pmatrix} U^{-1},
\eeq
then Claudson-Halpern-Samuel (CHS) solution, which has $N$ continuous parameters $k_i$, is given by
\beq
\psi(X)=\frac{1}{\cal M}\sum_{\sigma}\text{sgn}(\sigma)\exp\left[i\sum_{i=1}^Nk_{\sigma(i)}x_i\right],
\label{chss}
\eeq
where ${\cal M}=\prod_{i<j}(x_i-x_j)$ and $\sum_\sigma$ indicates summation over all the permutations.
This expression gives a solution in the case of U($N$) gauge group. To remove center of mass U(1) part
one just have to impose the condition $\sum_ik_i=0$. Then $\psi(X)$ depends on only traceless part
$w_i\equiv x_i-\frac{1}{N}\sum_jx_j$.
In SU(2) case this is the only energy eigenfunction with zero fermion number\cite{ch85,s07}.
In the case of higher groups there are more solutions \cite{trze06,k0910}.

Naively we can expect that the spectrum can be constructed by acting gauge invariant operators made of
$\Pi$ and $\theta^\dagger$ on states in the sector of zero fermion number\cite{s07}. Indeed in $N=2$ case
it has been proven that the whole spectrum is constructed in this way on CHS solution\cite{ch85,s07}.
To investigate the case of general $N$, in section 2 we compute the thermal partition function of this system,
and give an interpretation of the result. The form of the partition function gives support to 
the naive expectation, and we give a conjecture that CHS solution is the unique eigenfunction of
simultaneously diagonalizable operators $\text{tr}\;[\Pi^n]~(n=2,3,\dots, N)$.
In section 3, we prove this conjecture in $N=3$ and $N=4$ cases.
Section 4 contains some discussion, and 
Appendix contains necessary information on SU($N$) group theory.

\section{Thermal partition function and spectrum of $D=2$ matrix quantum mechanics}

In this section we compute the thermal partition function of $D=2$ quantum mechanics
$Z=\text{Tr}_{\text{phys}}\;[e^{-\beta H}]$, where the trace $\text{Tr}_{\text{phys}}$ is 
taken over the physical states. Then from it we will obtain insight to the structure of the spectrum of this system.
First we compute $Z$ in the operator formalism.
$Z$ can be rewritten in terms of a trace without the condition of gauge invariance, by inserting 
SU($N$) group element $e^{i\lambda^aG^a}$:
\beq
Z=\int d\lambda \text{Tr}\;[e^{-\beta h+i\lambda^aG^a}],
\eeq
where $\int d\lambda$ indicates integral over SU($N$) (See Appendix for notation about SU($N$) group theory used
here and the following),
and works as a projector onto the gauge invariant states. 
We can set $\lambda^{(ij)}=0$ without loss of generality. 

Since in $-\beta h+i\lambda^aG^a$ boson part and fermion part are decoupled,
$Z$ is decomposed into corresponding parts $Z_B$ and $Z_F$:
\beq
Z=\int d\lambda Z_B(\lambda)Z_F(\lambda),\quad
Z_B(\lambda)=\text{Tr}\;[e^{-\beta h+i\lambda^aG^a_B}],\quad Z_F(\lambda)=\text{Tr}\;[e^{i\lambda^a G^a_F}],
\eeq
where
\beq
G^a_B=f_{bac}X^b\Pi^c,\quad
G^a_F=-if_{bac}(\theta^b)^\dagger\theta^c.
\eeq

$Z_F$ can be computed by using the knowledge of group theory:
$e^{i\lambda^a G^a_F}$ is just the gauge group element for fermions:
\beq
e^{i\lambda^aG^a_F}(\theta_b)^\dagger e^{-i\lambda^a G^a_F}
=[\text{Ad}(\lambda)]_{bc}(\theta_c)^\dagger,
\eeq
and $(\theta^{a_1})^\dagger\dots(\theta^{a_n})^\dagger\ket{0_F}$ forms an antisymmetric tensor product representation
of adjoint representation. Therefore $Z_F$ is given by the sum of group theory characters of those representations:
\beq
Z_F(\lambda)=\sum_{n=0}^{N^2-1}\chi(\text{Alt}_n(\text{adj.}))=
2^{N-1}\prod_{\stackrel{\mbox{$\scriptstyle i,j=1$}}{\mbox{$\scriptstyle i\not\neq j$}}}^N(1+z_i/z_j),
\eeq
where we used \eqref{chaltadj}.
If we introduce an additional parameter $q$ for counting fermion number $F$:
$Z_F(\lambda,q)\equiv\text{Tr}\;[q^Fe^{i\lambda^a G^a_F}]$, then similarly,
\beq
Z_F(\lambda,q)=\sum_{n=0}^{N^2-1}q^n\chi(\text{Alt}_n(\text{adj.}))=(1+q)^{N-1}
\prod_{\stackrel{\mbox{$\scriptstyle i,j=1$}}{\mbox{$\scriptstyle i\not\neq j$}}}^N(1+qz_i/z_j).
\eeq

Next let us compute $Z_B(\lambda)$. By introducing eigenstates of $\Pi$: $\Pi^a\ket{p_b}=p^a\ket{p_b}$,
\bea
Z_B(\lambda) & = & \Big[\prod_a\int_{-\infty}^\infty \frac{dp_a}{2\pi}\Big]
\bra{p_b}e^{-\beta h}e^{i\lambda^c G^c_B}\ket{p_b}
\nn & = & 
\Big[\prod_a\int_{-\infty}^\infty\frac{dp_a}{2\pi}\Big]e^{-\frac{\beta g^2}{2}p_d^2}
\bra{p_b}e^{i\lambda^c G^c_B}\ket{p_b}.
\eea
Noting that $e^{i\lambda^c G^c_B}$ generates gauge group elements,
\beq
\bra{p_b}e^{i\lambda^c G^c_B}\ket{p_b}=\vev{p_b|\text{Ad}(\lambda)_{bc}p_c}
=\prod_b2\pi\delta(p_b-\text{Ad}(\lambda)_{bc}p_c).
\eeq
Since $\text{Ad}(\lambda)_{bc}$ is diagonalized by decomposing adjoint indices into Cartan part $m$ and 
ladder operator part $(ij)$,
\bea
\bra{p_b}e^{i\lambda^c G^c_B}\ket{p_b} & = & 
(2\pi\delta(0))^{N-1}\prod_{\stackrel{\mbox{$\scriptstyle i,j=1$}}{\mbox{$\scriptstyle i\not\neq j$}}}^N
2\pi\delta([1-z_i/z_j]p_{(ij)})
\nn & = & V^{N-1}(2\pi)^{N(N-1)}
\prod_{\stackrel{\mbox{$\scriptstyle i,j=1$}}{\mbox{$\scriptstyle i\not\neq j$}}}^N(1-z_i/z_j)^{-1}\delta(p_{(ij)}),
\eea
where $V$ is the volume factor $V=2\pi\delta(p=0)$. Therefore 
\bea
Z_B(\lambda) & = & \prod_{m=1}^{N-1}\Big[\int_{-\infty}^\infty \frac{dp_m}{2\pi}Ve^{-\frac{\beta g^2}{2}p_m^2}\Big]
\prod_{\stackrel{\mbox{$\scriptstyle i,j=1$}}{\mbox{$\scriptstyle i\not\neq j$}}}^N(1-z_i/z_j)^{-1}
\nn & = & 
\left(\frac{V}{\sqrt{2\pi\beta g^2}}\right)^{N-1}
\prod_{\stackrel{\mbox{$\scriptstyle i,j=1$}}{\mbox{$\scriptstyle i\not\neq j$}}}^N(1-z_i/z_j)^{-1}.
\eea
Then $Z$ is given by
\bea
Z & = & \frac{1}{N!}\prod_{i=1}^{N-1}\int_0^1 dy_i
\prod_{\stackrel{\mbox{$\scriptstyle i,j=1$}}{\mbox{$\scriptstyle i\not\neq j$}}}^N(z_i-z_j)
\nn & & 
\times 2^{N-1}\prod_{\stackrel{\mbox{$\scriptstyle i,j=1$}}{\mbox{$\scriptstyle i\not\neq j$}}}^N(1+z_i/z_j)
\left(\frac{V}{\sqrt{2\pi\beta g^2}}\right)^{N-1}
\prod_{\stackrel{\mbox{$\scriptstyle i,j=1$}}{\mbox{$\scriptstyle i\not\neq j$}}}^N(1-z_i/z_j)^{-1}
\nn & = & 
 \frac{1}{N!}\left(\frac{2V}{\sqrt{2\pi\beta g^2}}\right)^{N-1}
\prod_{i=1}^{N-1}\int_0^1 dy_i \Bigg[
\prod_{\stackrel{\mbox{$\scriptstyle i,j=1$}}{\mbox{$\scriptstyle i\not\neq j$}}}^N(1+z_i/z_j)\Bigg].
\eea
Note that the factor 
$\prod_{i\not\neq j}(1-z_i/z_j)^{-1}$
from $Z_B$ is canceled with the measure factor from $\int d\lambda$.

To confirm the above result from different viewpoint, we will quickly explain the path integral calculation of $Z$
(For similar calculation in a different context, see e.g. \cite{yin03}).
In this calculation the overall normalization of $Z$ is somewhat ambiguous, and therefore we shall not keep track of 
normalization factors.
In Euclidean path integral formulation $Z$ is given by
\beq
Z=\frac{1}{\text{Vol(SU($N$))}}\int\mathop{DA_0DXD\theta D\theta^\dagger}_{\text{periodicity cond.}}~
\exp\Bigg[-\frac{1}{g^2}\int_0^\beta dt~\text{tr}\,
\Big[\frac{1}{2}DXDX+ig^2\theta^\dagger D\theta-g^4\theta^\dagger[X,\theta]\Big]\Bigg],
\eeq
where $A_0(t)$ and $X(t)$ are periodic, and $\theta(t)$ is antiperiodic up to gauge transformation,
with periodicity $\beta$.
Vol(SU($N$)) is the infinite volume of the redundancy from the gauge symmetry.
By the shift $A_0\rightarrow A_0+g^2X$, we can eliminate the interaction term $\theta^\dagger[X,\theta]$:
\beq
Z=\frac{1}{\text{Vol(SU($N$))}}\int\mathop{DA_0DXD\theta D\theta^\dagger}_{\text{periodicity cond.}}~
\exp\Bigg[-\frac{1}{g^2}\int_0^\beta dt~\text{tr}\,
\Big[\frac{1}{2}DXDX+ig^2\theta^\dagger D\theta\Big]\Bigg].
\eeq
$A_0$ can be taken to be $\Delta$ given in the following, by taking Lorentz gauge $\p_tA_0=0$: 
\beq
\Delta=\begin{pmatrix} \alpha_1 & & & \\ & \alpha_2 & & \\ & & \ddots & \\ & & & \alpha_N \end{pmatrix},
\quad
\begin{matrix}
0\leq\alpha_1\leq\alpha_2\leq\dots\leq\alpha_{N-1}<2\pi/\beta, \\
\alpha_N=-(\alpha_1+\alpha_2+\dots+\alpha_{N-1}), \\
\alpha_i: t\text{-independent},
\end{matrix}
\eeq
General $A_0$ is in the form of gauge transformation of $\Delta$: 
$A_0(t)=U(t)\Delta U^{-1}(t)-i\p_t U(t)U^{-1}(t)$, and 
the path integral measure $DA_0$ is decomposed as follows:
\beq
DA_0=DU\Big[\prod_{i=1}^{N-1}d\alpha_i\Big]\text{Det}'[D],
\eeq
where Det${}'$ is the functional determinant for periodic functions with the zero eigenvalues removed.
Although we will see that the Faddeev-Popov determinant Det${}'[D]$ cancels, let us see how this is computed.
By decomposing the adjoint index $a$ into Cartan subalgebra part $m$ and ladder operator part $(ij)$,
\beq
\text{Det}'[D]=\prod_{m=1}^{N-1}\text{Det}'[\p_t]
\prod_{\stackrel{\mbox{$\scriptstyle i,j=1$}}{\mbox{$\scriptstyle i\not\neq j$}}}^N
\text{Det}[\p_t-i(\alpha_i-\alpha_j)].
\eeq
For periodic functions eigenvalues of $\p_t$ are given by $i\frac{2\pi}{\beta}n~(n\in\mathbb{Z})$
and 
\beq
\text{Det}'[\p_t]=\prod_{n\not\neq 0}i\frac{2\pi}{\beta}n=-i\beta,
\eeq
where we used the following zeta-function regularization:
\beq
\prod_{n=1}^\infty n^\alpha=(2\pi)^{\alpha/2},\quad \prod_{n=-\infty}^\infty a=1.
\eeq
Using the above and $\prod_{n=1}^\infty(1-\frac{x^2}{n^2})=\frac{\sin\pi x}{\pi x}$, 
$\prod_{i\not\neq j}\text{Det}[\p_t-i(\alpha_i-\alpha_j)]$ can be computed similarly, and then
\beq
\text{Det}'[D]\sim
\prod_{\stackrel{\mbox{$\scriptstyle i,j=1$}}{\mbox{$\scriptstyle i\not\neq j$}}}^N(z_i-z_j),
\eeq
where $z_i=e^{2\pi iy_i}$, and  $y_i=\frac{\beta}{2\pi}\alpha_i$. This is the measure factor
for the integration over SU($N$) group.
The path integral of $DU$ cancels the infinite volume factor: $\int DU=\text{Vol(SU($N$))}$.
Since $\alpha_i$ appear symmetrically, the integral of $\alpha_i$ can be done by regarding
all the $\alpha_i$ independent from each other, and dividing by the factor $1/N!$:
\beq
\frac{1}{N!}\prod_{i=1}^{N-1}\int_0^1dy_i.
\eeq
After the gauge fixing $X$ and $\theta$ can be taken as periodic and antiperiodic respectively.
The functional integral of them can be done as Gaussian integrals:
\bea
\int DX & \rightarrow & V^{N-1}\text{Det}'[D]^{-1},
\\
\int D\theta D\theta^\dagger & \rightarrow & 
\prod_{m=1}^{N-1}\wt{\text{Det}}\Big[\p_t\Big]
\prod_{\stackrel{\mbox{$\scriptstyle i,j=1$}}{\mbox{$\scriptstyle i\not\neq j$}}}^N
\wt{\text{Det}}\Big[\p_t-i(\alpha_i-\alpha_j)\Big]
 \sim
\prod_{\stackrel{\mbox{$\scriptstyle i,j=1$}}{\mbox{$\scriptstyle i\not\neq j$}}}^N(z_i+z_j),
\eea
where $\wt{\text{Det}}$ is the functional determinant for antiperiodic functions, and the volume factor $V^{N-1}$
comes from the integral of the zero modes of $X$. Then
\bea
Z & \sim & \frac{1}{N!}\prod_{i=1}^{N-1}\int_0^1dy_i~\text{Det}'[D]\cdot V^{N-1}\text{Det}'[D]^{-1}
\prod_{\stackrel{\mbox{$\scriptstyle i,j=1$}}{\mbox{$\scriptstyle i\not\neq j$}}}^N
(z_i+z_j)
\nn & = &
\frac{1}{N!}V^{N-1}\prod_{i=1}^{N-1}\int_0^1dy_i~\Bigg[
\prod_{\stackrel{\mbox{$\scriptstyle i,j=1$}}{\mbox{$\scriptstyle i\not\neq j$}}}^N(z_i+z_j)\Bigg].
\eea
Thus we can reproduce the result of operator formalism calculation, with the similar
cancellation of the measure factor.

To interpret the form of $Z$ and read off information on the spectrum,
let us consider mass deformation of this quantum mechanics given by the following action:
\bea
S & = & \frac{1}{g^2}\int dt\;\text{tr}\Big[
\frac{1}{2}DXDX-\frac{1}{2}g^4M^2X^2+ig^2\theta^\dagger D\theta-g^4M\theta^\dagger\theta
-g^4\theta^\dagger[X,\theta]
\Big].
\eea
The supersymmetry transformation is also deformed:
\beq
\delta X=i(\epsilon^\dagger\theta+\epsilon\theta^\dagger), \quad
\delta\theta=\left(iMX-\Pi\right)\epsilon, \quad
\delta A_0=g^2\delta X.
\eeq
Hamiltonian and supercharge are
\bea
H & = & g^2\text{tr}\;\Big[\frac{1}{2}\Pi^2+\frac{1}{2}M^2X^2+M\theta^\dagger\theta\Big]
\nn & &
+(\text{terms proportional to 1st class constraints}), \\
Q & = & -\text{tr}\;[(\Pi+iMX)\theta].
\eea
These satisfy the same commutation relations as those in the undeformed case.
Again $H$ acts as free Hamiltonian $h=g^2\text{tr}\;\big[
\frac{1}{2}\Pi^2+\frac{1}{2}M^2X^2+M\theta^\dagger\theta\big]$ on the physical states.
This system is easier than the undeformed case: Introducing annihilation operators
$A_a=\frac{1}{\sqrt{2M}}\Pi_a-i\sqrt{\frac{M}{2}}X_a$, $h$ is rewritten as
\beq
h=Mg^2(A^\dagger_aA_a+\theta^\dagger_a\theta_a).
\eeq
Therefore this system is equivalent to a supersymmetric harmonic oscillator system, 
with the physical condition $G\ket{\text{phys}}=0$.
The spectrum of states is constructed by acting $A_a^\dagger$ and $\theta_a^\dagger$ on the vacuum
$\ket{0}$ defined by $A_a\ket{0}=0$ and $\theta_a\ket{0}=0$.
For states in the following form:
\beq
A_{a_1}^\dagger\dots A_{a_n}^\dagger\theta_{b_1}^\dagger\dots\theta_{b_m}^\dagger\ket{0},
\eeq
indices $a_1,\dots, a_n$ can be symmetrized and $b_1,\dots, b_m$ antisymmetrized, and 
gauge invariant combinations of indices give physical states.
The number of such gauge invariant states can be computed by 
\beq
\int d\text{SU}(N) \chi(\text{Sym}_n(\text{adj.}))\chi(\text{Alt}_m(\text{adj.})).
\eeq
Therefore the partition function of this system $Z_M=\text{Tr}_{\text{phys}}[e^{-\beta H}]$
can be computed by noting that the factor $e^{-\beta H}$ gives $q=e^{-\beta g^2 M}$ for each $A_a^\dagger$
and $\theta_a^\dagger$ on $\ket{0}$:
\bea
Z_M & = & \int d\text{SU}(N)\sum_{n=0}^\infty q^n\chi(\text{Sym}_n(\text{adj.}))
\sum_{n=0}^\infty q^n\chi(\text{Alt}_n(\text{adj.}))
\nn & = & 
\frac{1}{N!}\int \prod_{i=1}^{N-1} dy_i
\Big[\prod_{\stackrel{\mbox{$\scriptstyle i,j=1$}}{\mbox{$\scriptstyle i\not\neq j$}}}^N(1-z_i/z_j)\Big]
\left(\frac{1+q}{1-q}\right)^{N-1}\prod_{\stackrel{\mbox{$\scriptstyle i,j=1$}}{\mbox{$\scriptstyle i\not\neq j$}}}^N
\frac{1+qz_i/z_j}{1-qz_i/z_j},
\label{udz}
\eea
where we used \eqref{chsymadj} and \eqref{chaltadj}. For some analysis of this expression see \cite{trze07-2}.

If we take massless limit $q\rightarrow 1$, we obtain 
almost the same expression of the partition function as that in the undeformed case: the factor 
$\prod_{i\not\neq j}(1-z_i/z_j)$ is canceled, and the factor $(1-q)^{1-N}$ corresponds to 
the infinite volume factor $V^{N-1}$. Keeping this in mind, let us return to
the structure of the spectrum in the undeformed case and try to understand the structure of $Z$ more.
If we have an energy eigenstate $\ket{E}$ with zero fermion number,
we obtain new energy eigenstates by acting gauge invariant operators made of $\Pi$ and $\theta^\dagger$ on $\ket{E}$,
because those operators commute with the Hamiltonian\cite{s07}.
Those operators can be expressed as products of single trace operators such as 
$\text{tr}\,[\Pi^n], \text{tr}\,[\Pi^n\theta^\dagger]$, or $\text{tr}\,[\Pi^n\theta^\dagger\Pi^m\theta^\dagger]$.
Note that at this stage we do not know if there are states which
cannot be constructed in this way from a solution with zero fermion number.

We can see a correspondence of states between mass deformed and undeformed systems:
\beq
A_{a_1}^\dagger\dots A_{a_n}^\dagger\theta_{b_1}^\dagger\dots\theta_{b_m}^\dagger\ket{0}
\quad \longleftrightarrow \quad 
\Pi_{a_1}\dots \Pi_{a_n}\theta_{b_1}^\dagger\dots\theta_{b_m}^\dagger\ket{E}.
\eeq
Therefore number of these states in the undeformed case can be counted by setting $q=1$ in \eqref{udz}.
However, due to the factor $(1-q)^{-1}$ we cannot take the limit $q\rightarrow 1$. This is not surprising, 
because there are infinitely many gauge invariant operators made of $\Pi$.
This problem can be avoided as follows:
Since $\text{tr}\,[\Pi^3], \text{tr}\,[\Pi^4], \dots$, and $\text{tr}\,[\Pi^N]$ are hermitian and commute
with each other, $h$ and $G$, these are simultaneously diagonalizable. (Note that in SU($N$) case $\text{tr}\,[\Pi^n]$
with $n>N$ can be expressed by products of those with lower $n$.) Therefore we consider an eigenstate
$\ket{p_2,p_3,\dots, p_N}$ defined by
\beq
\text{tr}\,[\Pi^n]\ket{p_2,p_3,\dots, p_N}=p_n\ket{p_2,p_3,\dots, p_N},\quad G^a\ket{p_2,p_3,\dots, p_N}=0.
\eeq
Since $\text{tr}\,[\Pi^n]$ do not produce new states when they act on $\ket{p_2,p_3,\dots, p_N}$,
we want to remove products of $\text{tr}\,[\Pi^n]$ from the list of the gauge invariant operators. 
The list of those operators can be given by expanding the following expression:
\bea
& & [1+q^2\text{tr}\,[\Pi^2]+(q^2\text{tr}\,[\Pi^2])^2+(q^2\text{tr}\,[\Pi^2])^3+\dots] \nn
& & \times[1+q^3\text{tr}\,[\Pi^3]+(q^3\text{tr}\,[\Pi^3])^2+(q^3\text{tr}\,[\Pi^3])^3+\dots] \nn
& & \dots \nn
& & \times[1+q^N\text{tr}\,[\Pi^N]+(q^N\text{tr}\,[\Pi^N])^2+(q^N\text{tr}\,[\Pi^N])^3+\dots] \nn
& & = \prod_{n=2}^N\sum_{m=0}^\infty(q^n\text{tr}\,[\Pi^n])^m,
\eea
where $q$ is introduced to count the number of $\Pi$. If one just want to count the number of these 
gauge invariant operators, $\text{tr}\,[\Pi^n]$ in the above expression can be replaced by 1: 
\beq
\prod_{n=2}^N\sum_{m=0}^\infty(q^n)^m=\prod_{n=2}^N(1-q^n)^{-1}
=(1-q)^{1-N}\prod_{n=2}^N(1+q+q^2+\dots+q^{n-1})^{-1}.
\eeq
The coefficient of $q^n$ in the expansion of this expression gives the number of the operators made of $n$ $\Pi$s.

Therefore the number of gauge invariant operators which act nontrivially on $\ket{p_2,p_3,\dots, p_N}$
can be counted by
\bea
& & \Big[\int d\text{SU}(N)\sum_{n=0}^\infty q^n\chi(\text{Sym}_n(\text{adj.}))
\sum_{n=0}^\infty q^n\chi(\text{Alt}_n(\text{adj.}))\Big]\Big/
\nn & & 
(1-q)^{1-N}\prod_{n=2}^N(1+q+q^2+\dots+q^{n-1})^{-1}
\nn
& = & 
\frac{1}{N!}\int_0^1 \prod_{i=1}^{N-1} dy_i
\Big[\prod_{\stackrel{\mbox{$\scriptstyle i,j=1$}}{\mbox{$\scriptstyle i\not\neq j$}}}^N(1-z_i/z_j)\Big]
(1+q)^{N-1}\prod_{n=2}^N(1+q+q^2+\dots+q^{n-1})
\nn & & \times
\prod_{\stackrel{\mbox{$\scriptstyle i,j=1$}}{\mbox{$\scriptstyle i\not\neq j$}}}^N\frac{1+qz_i/z_j}{1-qz_i/z_j}.
\eea
We can safely take the limit $q\rightarrow 1$ in this expression:
\beq
\int_0^1 \prod_{i=1}^{N-1}dy_i~2^{N-1}
\prod_{\stackrel{\mbox{$\scriptstyle i,j=1$}}{\mbox{$\scriptstyle i\not\neq j$}}}^N(1+z_i/z_j).
\label{udnum}
\eeq
Note that this expression is very close to $Z$.
\eqref{udnum} times the contribution from integrals of parameters $p_2, p_3,\dots,p_N$ gives
the contribution to the partition function from states constructed by acting the gauge invariant operators
on $\ket{p_2,p_3,\dots, p_N}$. At this stage we do not know how many independent
solutions with the same eigenvalues of $\text{tr}\,[\Pi^n]$ exist in the sector of zero fermion number.

CHS solution gives a hint to the contribution from integrals of parameters $p_2, p_3,\dots, p_N$. 
It has $N-1$ continuous parameters $k_1, k_2,\dots, k_{N-1}$ corresponding to free particle momenta,
and these appear symmetrically in the solution.
Therefore the contribution to the partition function is computed by regarding $k_i$ independent,
with the factor $1/N!$:
\beq
\frac{1}{N!}\left[\int\frac{dk}{2\pi}\bra{k}
e^{-\frac{1}{2}\beta g^2 k^2}\ket{k}\right]^{N-1}
=\frac{1}{N!}\left[V\int\frac{dk}{2\pi}e^{-\frac{1}{2}\beta g^2 k^2}\right]^{N-1}
=\frac{1}{N!}\left[\frac{V}{\sqrt{2\pi\beta g^2}}\right]^{N-1}.
\eeq
This times \eqref{udnum} gives exactly the same expression as $Z$:
\beq
\frac{1}{N!}\left[\frac{2V}{\sqrt{2\pi\beta g^2}}\right]^{N-1}\int_0^1 \prod_{i=1}^{N-1}dy_i~
\Big[\prod_{\stackrel{\mbox{$\scriptstyle i,j=1$}}{\mbox{$\scriptstyle i\not\neq j$}}}^N(1+z_i/z_j)\Big].
\eeq
Thus we have given evidence for the following naively conjectured structure of the spectrum of states:
\begin{quotation}
The whole spectrum of $D=2$ matrix quantum mechanics is constructed by acting 
gauge invariant operators made of $\Pi$ and at least one $\theta^\dagger$
on eigenstates of $\text{tr}\,[\Pi^n]~(n=2,3,\dots,N)$ in the sector of zero fermion number.
\end{quotation}
and if the normalization of $Z$ inferred from CHS solution is correct,
the following conjecture follows: 
\begin{quotation}
In the sector of zero fermion number,
CHS solution is the unique eigenfunction of $\text{tr}\,[\Pi^2], \text{tr}\,[\Pi^3],\dots$ and $\text{tr}\,[\Pi^N]$
for fixed eigenvalues.
\end{quotation}
Although we do not have proofs of this conjecture for general $N$,
for $N=2$ case this has been shown to be correct\cite{ch85,s07}.
In the next section we shall give proofs for $N=3$ and $N=4$ cases.

\section{Analysis of eigenfunctions of $\text{tr}\,[\Pi^n]$ in lower $N$ cases}

In this section we give proofs of the conjecture in the previous section
for $N=3$ and $N=4$ cases i.e.
we show that CHS solutions for any $N$ are eigenfunctions of $\text{tr}\,[\Pi^3]$ and $\text{tr}\,[\Pi^4]$,
and they are the unique eigenfunctions of those operators with zero fermion number
for fixed eigenvalues in $N=3$ and $N=4$ cases.

First we show that CHS solutions are eigenfunctions of $\text{tr}\,[\Pi^3]$ and $\text{tr}\,[\Pi^4]$.
Since the expression \eqref{chss} is regarded as the restriction of the gauge invariant solution to diagonal $X$,
we have to know how $\Pi$ acts on such restricted functions.
Note that on any function $f(X)$, $\Pi_{(ij)}$ can be replaced by $G_{(ij)}$ if $X$ is set
$Y\equiv\text{diag}(x_1,\dots,x_N)$\cite{wln89}.
\beq
\Pi_{(ij)}f(X)\Big|_{X=Y}=
 \frac{i}{x_i-x_j}G_{(ij)}f(X)\Big|_{X=Y},
\eeq
where $\Pi_a$ is regarded as $-i\frac{\p}{\p X_a}$.
Therefore, action of two or more $\Pi_{(ij)}$s on a gauge invariant function can be reduced to that of less $\Pi_{(ij)}$s
by replacing the first $\Pi_{(ij)}$ by $G_{(ij)}$, and moving it rightward using the following commutation relation
until it annihilates the function:
\beq
[G_{(ij)},\Pi_{(kl)}]=i\delta_{il}\delta_{jk}(\p_i-\p_j)+\delta_{il}\Pi_{(jk)}-\delta_{jk}\Pi_{(il)},
\eeq
where $\p_i=\frac{\p}{\p x_i}$ and $\Pi_{(ii)}=0$. These can be shown from \eqref{suncom} and $w_i=(\nu^i)_mX_m$.
$\Pi_m$ is also replaced by $\p_i$ with the following relation:
\beq
(\nu^i)_m\Pi_m=-i(\nu^i)_m\frac{\p}{\p X_m}=-i\sum_j(\nu^i\cdot\nu^j)\p_j=-i(\p_i-P),
\eeq
where $P$ is the center of mass momentum operator $P=\frac{1}{N}\sum_{i=1}^N\p_i$, and summation over $m$ is understood.
Since we set the eigenvalue of $P$
zero and $P$ commutes with any operator dependent only on $\p_i$ and the differences $x_i-x_j$,
we ignore terms proportional to $P$ in the following.

Thus we can reduce action of $\Pi$s on gauge invariant functions to $\p_i$.
Therefore we obtain
\bea
\text{tr}\,[\Pi^2]f(X)\Big|_{X=Y} & = & 
\Bigg[\sum_{i=1}^N(\nu^i)_{m_1}(\nu^i)_{m_2}\Pi_{m_1}\Pi_{m_2}
+\mathop{{\sum}'}_{i,j=1}^N\Pi_{(ij)}\Pi_{(ji)}\Bigg]f(X)\Big|_{X=Y}
\nn & = & 
-\left[\sum_{i=1}^N\p_i\p_i+\mathop{{\sum}'}_{i,j=1}^N\frac{2}{x_i-x_j}\p_i\right]f(X)\Big|_{X=Y},
\\
\text{tr}\,[\Pi^3]f(X)\Big|_{X=Y} & = & \Bigg[
\sum_{i=1}^N(\nu^i)_{m_1}(\nu^i)_{m_2}(\nu^i)_{m_3}\Pi_{m_1}\Pi_{m_2}\Pi_{m_3}
\nn & &
+3\mathop{{\sum}'}_{i,j=1}^N(\nu^i)_m\Pi_m\Pi_{(ji)}\Pi_{(ij)}
+\mathop{{\sum}'}_{i,j,k=1}^N\Pi_{(ji)}\Pi_{(kj)}\Pi_{(ik)}\Bigg]f(X)\Big|_{X=Y}
\nn & = & i\Bigg[
\sum_{i=1}^N(\p_i)^3+3\mathop{{\sum}'}_{i,j=1}^N\frac{3}{x_i-x_j}(\p_i)^2
\nn & &
+3\mathop{{\sum}'}_{i,j,k=1}^N\frac{1}{x_j-x_i}\frac{1}{x_k-x_i}\p_i\Bigg]
f(X)\Big|_{X=Y},
\\
\text{tr}\,[\Pi^4]f(X)\Big|_{X=Y} & = & \Bigg[
\sum_{i=1}^N(\nu^i)_{m_1}(\nu^i)_{m_2}(\nu^i)_{m_3}(\nu^i)_{m_4}\Pi_{m_1}\Pi_{m_2}\Pi_{m_3}\Pi_{m_4}
\nn & &
+4\mathop{{\sum}'}_{i,j=1}^N(\nu^j)_{m_1}(\nu^j)_{m_2}\Pi_{m_1}\Pi_{m_2}\Pi_{(ji)}\Pi_{(ij)}
\nn & &
+2\mathop{{\sum}'}_{i,j=1}^N(\nu^j)_{m_1}(\nu^i)_{m_2}\Pi_{m_1}\Pi_{m_2}\Pi_{(ji)}\Pi_{(ij)}
\nn & &
+4\mathop{{\sum}'}_{i,j,k=1}^N(\nu^j)_m\Pi_m\Pi_{(ji)}\Pi_{(kj)}\Pi_{(ik)}
\nn & &
+\mathop{{\sum}'}_{i,j=1}^N\Pi_{(ji)}\Pi_{(ij)}\Pi_{(ji)}\Pi_{(ij)}
+2\mathop{{\sum}'}_{i,j,k=1}^N\Pi_{(ji)}\Pi_{(ij)}\Pi_{(kj)}\Pi_{(jk)}
\nn & &
+\mathop{{\sum}'}_{i,j,k,l=1}^N\Pi_{(ji)}\Pi_{(kj)}\Pi_{(lk)}\Pi_{(il)}
\Bigg]f(X)\Big|_{X=Y}
\nn & = & \Bigg[
\sum_{i=1}^N(\p_i)^4+\mathop{{\sum}'}_{i,j=1}^N\frac{4}{x_i-x_j}(\p_i)^3
\nn & &
+6\mathop{{\sum}'}_{i,j,k=1}^N\frac{1}{x_j-x_i}\frac{1}{x_k-x_i}(\p_i)^2
\nn & &
-4\mathop{{\sum}'}_{i,j,k,l=1}^N\frac{1}{x_j-x_i}\frac{1}{x_k-x_i}\frac{1}{x_l-x_j}\p_i
\Bigg]f(X)\Big|_{X=Y},
\eea
where $\sum'$ indicates summations with the condition that the dummy indices take different values from each other.
CHS solutions have the prefactor ${\cal M}$. This can be moved leftward using
\beq
\p_i\frac{1}{\cal M}=\frac{1}{\cal M}
\Big(\p_i+\sum_{\stackrel{\mbox{$\scriptstyle j=1$}}{\mbox{$\scriptstyle j\not\neq i$}}}^N\frac{1}{x_j-x_i}\Big),
\eeq
and with the following identities:
\bea
0 & = & \mathop{{\sum}'}_{i,j,k=1}^N\frac{1}{(x_j-x_i)(x_k-x_i)},
\nn
0 & = & \mathop{{\sum}'}_{i,j,k,l=1}^N\frac{1}{(x_j-x_i)(x_k-x_i)(x_l-x_i)},
\nn
0 & = & \mathop{{\sum}'}_{i,j,k,l,p=1}^N\frac{1}{(x_j-x_i)(x_k-x_i)(x_l-x_i)(x_p-x_i)},
\eea
we can show
\bea
\text{tr}\,[\Pi^2]\frac{1}{\cal M}f(X)\Big|_{X=Y} & = & -\frac{1}{\cal M}
\sum_{i=1}^N(\p_i)^2f(X)\Big|_{X=Y},
\label{mp2} \\
\text{tr}\,[\Pi^3]\frac{1}{\cal M}f(X)\Big|_{X=Y} & = & \frac{i}{\cal M}
\sum_{i=1}^N(\p_i)^3f(X)\Big|_{X=Y},
\\
\text{tr}\,[\Pi^4]\frac{1}{\cal M}f(X)\Big|_{X=Y} & = & \frac{1}{\cal M}
\sum_{i=1}^N(\p_i)^4f(X)\Big|_{X=Y}.
\label{mp4}
\eea
Obviously $\sum_{\sigma}\text{sgn}(\sigma)\exp\left[i\sum_{i=1}^Nk_{\sigma(i)}x_i\right]$
are eigenfunctions of $\sum_{i=1}^N(\p_i)^3$ and $\sum_{i=1}^N(\p_i)^4$, and therefore
the above equations mean that CHS solutions are eigenfunctions of $(\Pi^3)$ and $(\Pi^4)$.

Next, using the method in \cite{trze06}, we shall show that
eigenfunctions of $\text{tr}\,[\Pi^n]$ with zero fermion number for fixed eigenvalues
are unique in $N=3$ and $N=4$ cases.
Gauge invariant states in the sector of zero fermion number are in the following form:
\beq
\psi(X)=\sum_{n_2,n_3,\dots, n_N=0}^\infty a_{n_2,n_3,\dots,n_N}(X^2)^{n_2}(X^3)^{n_3}\dots(X^N)^{n_N},
\eeq
where $(A)=\text{tr}\,[A]$. The following expression in $(\Pi^n)\psi$, 
\beq
(\Pi^n)[(X^2)^{n_2}(X^3)^{n_3}\dots(X^N)^{n_N}]
\eeq
can be rewritten as a sum of terms in the form of
$(X^2)^{m_2}(X^3)^{m_3}\dots(X^N)^{m_N}$. Therefore eigenvalue equations $(\Pi^n)\psi=p_n\psi$ give
recurrence relations for determining $a_{n_2,n_3,\dots,n_N}$. If these relations determine them uniquely
up to the overall scaling, the eigenfunction must be unique.

For $N=2$, the recurrence relation contains $a_n$ and $a_{n+1}$, which determines $a_n$ 
completely in terms of $a_0$ \cite{trze06}.

For $N=3$, case,
by explicit calculation using \eqref{tr1}, \eqref{tr2} and the following equations:
\bea
{}[(\Pi^2), (X^2)^n] & = & 2n(3-2n-N^2)(X^2)^{n-1}-4ni(X^2)^{n-1}(X\Pi), \\
{}[(\Pi^2), (X^3)^n] & = & 9n(n-1)\left\{\frac{1}{N}(X^2)^2-(X^4)\right\}(X^3)^{n-2}
-6ni(X^3)^{n-1}(X^2\Pi), \\
{}[(X\Pi), (X^3)^n] & = & -3ni(X^3)^n, \\
{}[(\Pi^3), (X^2)^n] & = & 8in(n-1)(n-2)(X^3)(X^2)^{n-3}
\nn & & 
-6ni(X^2)^{n-1}(X\Pi^2)-12n(n-1)(X^2)^{n-2}(X^2\Pi), \\
{}[(\Pi^3), (X^3)^n] & = & i\left\{3n\left(\frac{4}{N}-5N+N^3\right)
-27n(n-1)\left(\frac{4}{N}-N\right)\right\}(X^3)^{n-1}
\nn & & 
+27in(n-1)(n-2)\left\{\frac{2}{N^2}(X^2)^3-\frac{3}{N}(X^2)(X^4)+(X^6)\right\}(X^3)^{n-3}
\nn & & 
-9ni(X^3)^{n-1}(X^2\Pi^2)+\frac{9n}{N}i(X^2)(X^3)^{n-1}(\Pi^2)
\nn & & 
+9n\left(\frac{4}{N}-N\right)(X^3)^{n-1}(X\Pi)-27n(n-1)(X^3)^{n-2}(X^4\Pi)
\nn & & 
+\frac{54}{N}n(n-1)(X^2)(X^3)^{n-2}(X^2\Pi),
\\
{}[(X\Pi^2), (X^3)^n] & = & \left\{3n\left(\frac{4}{N}-N\right)
+\frac{18}{N}n(n-1)\right\}(X^2)(X^3)^{n-1}-9n(n-1)(X^5)(X^3)^{n-2}
\nn & & 
-6ni(X^3)^{n-1}(X^3\Pi)+\frac{6n}{N}i(X^2)(X^3)^{n-1}(X\Pi),
\\
{}[(X^2\Pi), (X^3)^n] & = & -3ni(X^4)(X^3)^{n-1}+\frac{3n}{N}i(X^2)^2(X^3)^{n-1},
\eea
and the following equations which hold in $N=3$ case,
\beq
(X^4)=\frac{1}{2}(X^2)^2,\quad (X^5)=\frac{5}{6}(X^2)(X^3),\quad (X^6)=\frac{1}{4}(X^2)^3+\frac{1}{3}(X^3)^2,
\eeq
we obtain two recurrence relations, respectively from $(\Pi^2)\psi=p_2\psi$ and $(\Pi^3)\psi=p_3\psi$:
\bea
p_2a_{n,m} & = & -4(n+1)(n+3m+4)a_{n+1,m}-\frac{3}{2}(m+1)(m+2)a_{n-2,m+2},
\label{rrsu3p2}\\
-ip_3a_{n,m} & = & (m+1)\{30n+9nm+6n(n-1)+9m(m+4)+40\}a_{n,m+1}
\nn & & 
-\frac{3}{4}(m+1)(m+2)(m+3)a_{n-3,m+3}
\nn & &
+8(n+1)(n+2)(n+3)a_{n+3,m-1},
\label{rrsu3p3}
\eea
where $n, m\geq 0$ and terms proportional to $a_{n',m'}$ with negative $n'$ or $m'$ must be omitted. 
From \eqref{rrsu3p2} for $n=0$ and 1,
\beq
a_{1,m}=-\frac{p_2}{4(3m+4)}a_{0,m},\quad 
a_{2,m}=-\frac{p_2}{8(3m+5)}a_{1,m}=\frac{(p_2)^2}{32(3m+4)(3m+5)}a_{0,m},
\label{rrsu3p2-3}
\eeq
which means that $a_{1,m}$ and $a_{2,m}$ are determined by $a_{0,m}$.
Furthermore by the following relation from \eqref{rrsu3p2} for $n\geq 3$:
\beq
a_{n,m}=-\frac{1}{4n(n+3m+3)}\left\{p_2a_{n-1,m}+\frac{3}{2}(m+1)(m+2)a_{n-3,m+2}\right\},
\label{rrsu3p2-2}
\eeq
$a_{n,m}$ with $n\geq 3$ can also be determined recursively from $a_{0,m}$. Therefore there are 
infinitely many solutions to $(\Pi^2)\psi=p_2\psi$ and are determined by giving $a_{0,m}$ as an initial condition.

Next we shall analyze \eqref{rrsu3p3}. For $n=0$ and $m\geq 1$,
\beq
-ip_3a_{0,m} = (m+1)\{9m(m+4)+40\}a_{0,m+1}+48a_{3,m-1}.
\eeq
Eliminating $a_{3,m-1}$ in this equation by using \eqref{rrsu3p2-2},
\beq
-ip_3a_{0,m} = (m+1)\{9m(m+4)+40\}a_{0,m+1}
-\frac{4}{3m+3}\left\{p_2a_{2,m-1}+\frac{3}{2}m(m+1)a_{0,m+1}\right\},
\eeq
and eliminating $a_{2,m-1}$ again in this equation by using \eqref{rrsu3p2-3},
\beq
0=\big[(m+1)\{9m(m+4)+40\}-2m\big] a_{0,m+1}+ip_3a_{0,m}-\frac{(p_2)^3}{8(3m+1)(3m+2)(3m+3)}a_{0,m-1}.
\label{su3frr}
\eeq
This means that $a_{0,m}$ with $m\geq 2$ are determined from $a_{0,0}$ and $a_{0,1}$.
Again from \eqref{rrsu3p3} for $m=0$ and $n=0,1,2$,
\beq
a_{0,1}=-\frac{ip_3}{40}a_{0,0},\quad
a_{1,1}=-\frac{ip_3}{70}a_{1,0},\quad
a_{2,1}=-\frac{ip_3}{112}a_{2,0}.
\eeq
The first equation of the above means that the entire solution is determined by $a_{0,0}$ and shows that the
solution for fixed $p_2$ and $p_3$ is unique. The second and third ones are consistent with \eqref{rrsu3p2-3}.
Since it seems difficult to give simple expression of the solution to \eqref{su3frr}, we only show some of $a_{n,m}$: 
\beq
a_{0,2}=\frac{p_2^3 - 24 p_3^2}{161280}a_{0,0},\quad a_{0,3}=-\frac{ip_3(p_2^3-12p_3^2)}{35481600}a_{0,0},\quad
a_{0,4}=\frac{p_2^6-72p_2^3p_3^2+576p_3^4}{1549836288000}a_{0,0},\quad\dots
\eeq
This solution must coincide with CHS solution. Indeed by straightforward calculation
we have checked that lower terms in the expansion of CHS solution reproduce the above $a_{n,m}$.

Thus we see that for $N=3$ our conjecture given in the previous section is correct.
Next we shall analyze $N=4$ case. By the calculation similar to $N=3$ case we obtain three recurrence relations:
\bea
p_2a_{n,m,l} & = & 2(l+1)(l+2)a_{n-3,m,l+2}+ 
 \frac{9}{4}(m+1)(m+2)a_{n-2,m+2,l}
\nn & &
-(l+1)(14 m+12 1+29)a_{n-1,m,l+1}- 
 \frac{4}{3}(l+1)(l+2)a_{n,m-2,l+2}
\nn & &
-9(m+1)(m+2)a_{n,m+2,l-1}- 
 2(n+1)(15+2n+6m+8l))a_{n+1,m,l},
\label{su4p2}
\\
-ip_3a_{n,m,l} & = & 
8(n+1)(n+2)(n+3)a_{n+3,m-1,l}+
36(n+1)(n+2)(m+1)a_{n+2,m+1,l-1}
\nn & &
+4 (n+1)(l+1)(12 + 2 l + 3 m + 7 n)a_{n+1,m-1,l+1}
\nn & & 
+9(m+1)(15 + 18 l + 4 l^2 + 8 m + 4 l m + m^2 + 7 n + 8 l n + 2 m n - n^2)a_{n,m+1,l}
\nn & & 
-\frac{8}{9}(l+1)(l+2)(l+3)a_{n,m-3,l+3}
\nn & & 
+(l+1)(l+2)(32 + 28 n + 10 m + 12 l)a_{n-1,m-1,l+2}
\nn & & 
+9(m+1)(l+1)(3/2 + l - 2 n)a_{n-2,m+1,l+1}
\nn & & 
+6(l+1)(l+2)(l+3)a_{n-3,m-1,l+3}-
\frac{9}{2}(m+1)(l+1)(l+2)a_{n-4,m+1,l+2},
\label{su4p3} \\
p_4 a_{n,m,l} & = & 
\frac{27}{4}(m+1)(m+2)(l+1)(l+2)a_{n-5,m+2,l+2}
\nn & & 
-(l+1)(l+2)(l+3)(37 + 12 l + 26 m + 8 n) a_{n-4,m,l+3}
\nn & & 
+\frac{81}{64} (m+1)(m+2)(m+3)(m+4)a_{n-4,m+4,l}
\nn & & 
+\frac{28}{3}(l+1)(l+2)(l+3)(l+4)a_{n-3,m-2,l+4}
\nn & & 
-\frac{135}{8}(m+1)(m+2)(l+1)(9 + 4 l + 2 m) a_{n-3,m+2,l+1}
\nn & & 
+\frac{1}{4}(l+1)(l+2)(1791 + 1240 l + 192 l^2 + 1878m + 688 lm 
\nn & & 
+ 318m^2 + 24n - 48mn - 96n^2) a_{n-2,m,l+2}
\nn & & 
-\frac{81}{8} (m+1)(m+2)(m+3)(m+4)a_{n-2,m+4,l-1}
\nn & & 
+\frac{2}{9}(l+1)(l+2)(l+3)(69 + 36 l + 50 m + 136 n) a_{n-1,m-2,l+3}
\nn & & 
+\frac{9}{4} (m+1)(m+2)(117 + 222 l + 72 l^2 + 50m + 60 lm 
\nn & & 
+ 4m^2 - 51n - 24 ln - 14mn - 6n^2) a_{n-1,m+2,l}
\nn & & 
+\frac{28}{27}(l+1)(l+2)(l+3)(l+4)a_{n,m-4,l+4}
\nn & & 
+\frac{1}{2}(l+1)(1335 + 1384 l + 384 l^2 + 32 l^3 + 1284m + 764 lm + 64 l^2m 
\nn & & 
+ 360m^2 + 84 lm^2 + 24m^3 + 2102n + 1768 ln + 320 l^2n + 1944mn 
\nn & & 
+ 816 lmn + 336m^2n + 324n^2 + 192 ln^2 + 168mn^2 - 32n^3) a_{n,m,l+1}
\nn & & 
+\frac{81}{4} (m+1)(m+2)(m+3)(m+4)a_{n,m+4,l-2}
\nn & & 
+\frac{4}{9}(l+1)(l+2)(n+1)(135 + 32 l + 18m + 78n) a_{n+1,m-2,l+2}
\nn & & 
+ 9 (m+1)(m+2)(n+1)(45 + 24 l + 14m + 6n) a_{n+1,m+2,l-1}
\nn & & 
+(n+1)(n+2)(435 + 480 l + 96 l^2 + 396m + 192 lm 
\nn & & 
+ 72m^2 + 116n + 96 ln + 56mn) a_{n+2,m,l}
\nn & & 
+\frac{32}{3}(l+1)(n+1)(n+2)(n+3)a_{n+3,m-2,l+1}
\nn & & 
+16 (n+1)(n+2)(n+3)(n+4)a_{n+4,m,l-1},
\label{su4p4}
\eea
where $n, m,l\geq 0$ and terms proportional to $a_{n',m',l'}$ with negative $n'$, $m'$ or $l'$ must be omitted. 
From \eqref{su4p2}, we obtain the following equation which determines $a_{n+1,m,l}$ with $n\geq 0$,
$m\geq 0$, and $l\geq 0$, in terms of $a_{n',m',l'}$ with $n'\leq n$:
\bea
a_{n+1,m,l} & = & \frac{1}{2(n+1)(2n+6m+8l+15)}\Big[2(l+1)(l+2)a_{n-3,m,l+2} \nn
& & +\frac{9}{4}(m+1)(m+2)a_{n-2,m+2,l} - (l+1)(14m+12l+29)a_{n-1,m,l+1} \nn
& & - \frac{4}{3}(l+1)(l+2)a_{n,m-2,l+2} - p_2a_{n,m,l} - 9(m+1)(m+2)a_{n,m+2,l-1}\Big].
\eea
Therefore we only have to determine $a_{0,m,l}$. By setting $n=0$ and $l\geq 3$ in \eqref{su4p3}
and using the above to rewrite $a_{n,m,l}$ with $n\geq 1$ in terms of $a_{0,m,l}$, 
we obtain the following equation, which
determines $a_{0,m+5,l}$ with $m\geq 0$ and $l\geq 0$ in terms of $a_{0,m',l'}$ with $m'\leq 4$:
\bea
a_{0,m+5,l} & = &  \frac{1}{(m+1)(m+2)(m+3)(m+4)(m+5)(37+5m+8 l)}\Big[
\nn & &
\frac{64}{19683}(l+4)(l+5)(l+6)(l+7)(l+8)(l+9)a_{0,m-7,l+9}
\nn & &
+\frac{16}{2187}\frac{(l+4)(l+5)(l+6)(l+7)(39+6m+8 l)}{35+6m+ 8 l}p_2a_{0,m-5,l+7}
\nn & &
+\frac{4}{729}\frac{(l+4)(l+5)(41+6m+8 l)}{33+6m+8 l} p_2^2a_{0,m-3,l+5}
\nn & &
+\frac{8}{6561}(l+4)(l+5)(l+6) (119064 + 69256 l + 13584 l^2 + 896 l^3
\nn & &
 + 53865 m + 21600 l m + 2160 l^2 m + 9450 m^2 + 1890 l m^2 + 540 m^3)a_{0,m-3,l+6}
\nn & &
+\frac{1}{729}\frac{(39+6m+8 l)(41+6m+8 l)}{(33+6m+8 l) (35+6m+8 l)}p_2^3a_{0,m-1,l+3}
\nn & &
+\frac{2}{729}\frac{(l+4)(39+6m+8 l)}{35+6m+8 l}(16569 + 9730 l + 1920 l^2 + 128 l^3 
\nn & &
+ 7497 m + 2952 l m + 288 l^2 m + 
      1620 m^2 + 324 l m^2 + 108 m^3) p_2a_{0,m-1,l+4}
\nn & & 
-\frac{i}{729}(37+6m+8 l)(39+6m+8 l)(41+6m+8 l)p_3a_{0,m,l+3}
\nn & &
-\frac{1}{81}\frac{(m+1)(33+3m+8 l)(41+6m+8 l)}{33+6m+8 l} p_2^2a_{0,m+1,l+2}
\nn & &
- \frac{1}{81}(m+1) (5618265 + 5767832 l + 2369580 l^2 + 486560 l^3 + 
    49920 l^4 + 2048 l^5
\nn & &
 + 3603906 m + 2956578 l m + 911180 l^2 m + 124800 l^3 m + 6400 l^4 m
\nn & &
 + 926103 m^2 + 567682 l m^2 + 116580 l^2 m^2 + 8000 l^3 m^2 + 117882 m^3
\nn & &
 + 47628 l m^3 + 4860 l^2 m^3 + 7668 m^4 + 1512 l m^4 + 216 m^5)a_{0,m+1,l+3}
\nn & &
-\frac{1}{9}\frac{(m+1)(m+2)(m+3)(39+6m+8 l)(70+9m+16 l)}{35+6m+8 l}p_2a_{0,m+3,l+1}
\Big].
\eea
For later use we give some more equations given by setting 
$(n,m,l)=$
$(0,0,2)$,
$(0,0,1)$, $(0,3,0)$, $(0,2,0)$, $(0,1,0)$, $(0,0,0)$ in \eqref{su4p3}:
\bea
-ip_3a_{0, 0, 2} & = & 
 \frac{9}{899}p_2^2a_{0, 1, 1} + \frac{184275}{341}a_{0, 1, 2} + \frac{324}{319}p_2a_{0, 3, 0},
\label{eeq1} \\
-ip_3a_{0, 1, 1} & = & -\frac{1}{15525}p_2^3a_{0, 0, 1} - \frac{2024}{1395}p_2a_{0, 0, 2} +
 \frac{40}{2001}p_2^2a_{0, 2, 0} + \frac{720720}{899}a_{0, 2, 1},
\\
-ip_3a_{0, 0, 1} & = & \frac{3}{161}p_2^2a_{0, 1, 0} + \frac{6885}{23}a_{0, 1, 1},
\\
-ip_3a_{0, 3, 0} & = & -\frac{8}{27621}p_2^2a_{0, 0, 2} - 
  \frac{4240}{341}a_{0, 0, 3} - \frac{1}{24273}p_2^3a_{0, 2, 0}
\nn & &
- \frac{780}{899}p_2a_{0, 2, 1} + \frac{555120}{341}a_{0, 4, 0},
\\
-ip_3a_{0, 2, 0} & = & -\frac{1}{12075}p_2^3a_{0, 1, 0} - 
  \frac{570}{667}p_2a_{0, 1, 1} + \frac{129789}{145}a_{0, 3, 0},
\\
-ip_3a_{0, 1, 0} & = & -\frac{1}{4845}p_2^3a_{0, 0, 0} - 
  \frac{336}{391}p_2a_{0, 0, 1} + \frac{181440}{437}a_{0, 2, 0},
\\
-ip_3a_{0, 0, 0} & = & 135a_{0, 1, 0}.
\label{eeq7}
\eea
Henceforth we only show schematic forms of equations using the following symbol:
\beq
[n,m,l]\equiv\text{a term proportional to}~a_{n,m,l},
\eeq
because explicit expressions are lengthy and it is not illuminating to show details of them.
However we have computed all the explicit expressions and have confirmed that there is no singular coefficient
which prevents us from solving linear equations.
Setting $n=0$ in \eqref{su4p4}, rewriting $a_{n,m,l}$ with $n\geq 1$ in terms of $a_{0,m,l}$, and
setting $m=0,1,2,3,4$, we obtain
\bea
[0,0,l] & = & [0,0,l-1]+[0,0,l]+[0,0,l+1]+[0,2,l-2]+[0,2,l-1]
\nn & &
+[0,4,l-3]+[0,4,l-2]+[0,6,l-4]+[0,8,l-5],
\label{a001} \\
{}[0,1,l] & = & [0,1,l-1]+[0,1,l]+[0,1,l+1]+[0,3,l-2]+[0,3,l-1]
\nn & &
+[0,5,l-3]+[0,5,l-2]+[0,7,l-4]+[0,9,l-5],
\label{a011} \\
{}[0,2,l] & = & [0,0,l+1]+[0,0,l+2]+[0,2,l-1]+[0,2,l]+[0,2,l+1]+[0,4,l-2]
\nn & &
+[0,4,l-1]+[0,6,l-3]+[0,6,l-2]+[0,8,l-4]+[0,10,l-5],
\label{a021} \\
{}[0,3,l] & = & [0,1,l+1]+[0,1,l+2]+[0,3,l-1]+[0,3,l]+[0,3,l+1]+[0,5,l-2]
\nn & &
+[0,5,l-1]+[0,7,l-3]+[0,7,l-2]+[0,9,l-4]+[0,11,l-5],
\label{a031} \\
{}[0,4, l] & = & [0,0,l+3]+[0,0,l+4]+[0,2,l+1]+[0,2,l+2]
\nn & &
+[0,4,l-1]+[0,4,l]+[0,4,l+1]+[0,6,l-2]+[0,6,l-1]
\nn & &
+[0,8,l-3]+[0,8,l-2]+[0,10,l-4]+[0,12,l-5].
\label{a041}
\eea
Rewriting $[0,m,l]$ with $m\geq 5$ further in terms of $[0,m',l']$ with $m'=0,1,2,3,4$, 
we obtain the following: from \eqref{a001}, 
\bea
[0,0,0] & = & [0,0,0]+[0,0,1], \label{a000-1} \\
{}[0,0,1] & = & [0,0,0]+[0,0,1]+[0,0,2]+[0,2,0], \label{a000-2}\\
{}[0,0,2] & = & [0,0,1]+[0,0,2]+[0,0,3]+[0,2,0]+[0,2,1]+[0,4,0], \label{a000-3}\\
{}[0,0,3] & = & [0,0,2]+[0,0,3]+[0,0,4]+[0,2,1]+[0,2,2]+[0,4,0]+[0,4,1], \\
{}[0,0,4] & = & [0,0,3]+[0,0,4]+[0,0,5]+[0,1,3]+[0,2,2]+[0,2,3]
\nn & &
+[0,4,1]+[0,4,2], \\
{}[0,0,l] & = & [0,0,l-1]+[0,0,l]+[0,0,l+1]+[0,1,l-1]+[0,2,l-2]
\nn & &
+[0,2,l-1]+[0,3,l-2]+[0,4,l-3]+[0,4,l-2], \label{a000-5}
\eea
where $l\geq 5$. From \eqref{a011},
\bea
[0,1,0] & = & [0,1,0]+[0,1,1], \label{a010-1}\\
{}[0,1,1] & = & [0,1,0]+[0,1,1]+[0,1,2]+[0,3,0], \\
{}[0,1,2] & = & [0,0,3]+[0,1,1]+[0,1,2]+[0,1,3]+[0,3,0]+[0,3,1], \\
{}[0,1,3] & = & [0,0,3]+[0,0,4]+[0,1,2]+[0,1,3]+[0,1,4]+[0,3,1]+[0,3,2], \\
{}[0,1,4] & = & [0,0,4]+[0,0,5]+[0,1,3]+[0,1,4]+[0,1,5]+[0,2,3]
\nn & &
+[0,3,2]+[0,3,3], \label{a010-5} \\
{}[0,1,l] & = & [0,0,l]+[0,0,l+1]+[0,1,l]+[0,1,l+1]+[0,2,l-1]
\nn & &
+[0,3,l-1]+[0,4,l-2],
\eea
where $l\geq 5$. From \eqref{a021},
\bea
[0, 2, 0] & = & [0, 0, 1]+[0, 0, 2]+[0, 2, 0]+[0, 2, 1], \label{a020-1}\\
{}[0, 2, 1] & = & [0, 0, 2]+[0, 0, 3]+[0, 2, 0]+[0, 2, 1]+[0, 2, 2]+[0, 4, 0], \\
{}[0, 2, 2] & = & [0, 0, 3]+[0, 0, 4]+[0, 1, 3]+[0, 2, 1]+[0, 2, 2]+[0, 2, 3]
\nn & &
+[0, 4, 0]+[0, 4, 1], \\
{}[0, 2, 3] & = & [0, 0, 3]+[0, 0, 4]+[0, 0, 5]+[0, 1, 3]+[0, 1, 4]
\nn & &
+[0, 2, 2]+[0, 2, 3]+[0, 2, 4]+[0, 4, 1]+[0, 4, 2],
\\
{}[0, 2, 4] & = &[0, 0, 4]+[0, 0, 5]+[0, 0, 6]+[0, 1, 4]+[0, 1, 5]
\nn & &
+[0, 2, 3]+[0, 2, 4]+[0, 2, 5]+[0, 3, 3]+[0, 4, 2]+[0, 4, 3], \\
{}[0,2,l] & = & [0,0,l]+[0,0,l+1]+[0,0,l+2]+[0,1,l]+[0,2,l-1]
\nn & &
+[0,2,l]+[0,2,l+1]+[0,3,l-1]+[0,4,l-2]+[0,4,l-1], \label{a020-5}
\eea
where $l\geq 5$. From \eqref{a031},
\bea
[0, 3, 0] & = & [0, 1, 1]+[0, 1, 2]+[0, 3, 0]+[0, 3, 1], \label{a030-1}\\
{}[0, 3, 1] & = & [0, 0, 3]+[0, 1, 2]+[0, 1, 3]+[0, 3, 0]+[0, 3, 1]+[0, 3, 2], \\
{}[0, 3, 2] & = & [0, 0, 3]+[0, 0, 4]+[0, 1, 2]+[0, 1, 3]+[0, 1, 4]+[0, 2, 3]
\nn & &
+[0, 3, 1]+[0, 3, 2]+[0, 3, 3], \\
{}[0, 3, 3] & = & [0, 0, 4]+[0, 0, 5]+[0, 1, 3]+[0, 1, 4]+[0, 1, 5]
\nn & &
+[0, 2, 3]+[0, 2, 4]+[0, 3, 2]+[0, 3, 3]+[0, 3, 4], \\
{}[0, 3, 4] & = & [0, 0, 5]+[0, 0, 6]+[0, 1, 4]+[0, 1, 5]+[0, 1, 6]
\nn & &
+[0, 2, 4]+[0, 2, 5]+[0, 3, 3]+[0, 3, 4]+[0, 3, 5]+[0, 4, 3], \\
{}[0,3,l] & = & [0,0,l+1]+[0,0,l+2]+[0,1,l+1]
\nn & &
+[0,1,l+2]+[0,2,l]+[0,2,l+1]+[0,3,l]
\nn & &
+[0,3,l+1]+[0,4,l-1], \label{a030-5}
\eea
where $l\geq 5$. From \eqref{a041},
\bea
[0, 4, 0] & = & [0, 0, 3]+[0, 0, 4]+[0, 2, 1]+[0, 2, 2]+[0, 4, 0]+[0, 4, 1], \label{a040-1}\\
{}[0, 4, 1] & = & [0, 0, 3]+[0, 0, 4]+[0, 0, 5]+[0, 1, 3]+[0, 2, 2]+[0, 2, 3]
\nn & &
+[0, 4, 0]+[0, 4, 1]+[0, 4, 2], \\
{}[0, 4, 2] & = & [0, 0, 3]+[0, 0, 4]+[0, 0, 5]+[0, 0, 6]+[0, 1, 3]+[0, 1, 4]
\nn & &
+[0, 2, 2]+[0, 2, 3]+[0, 2, 4]+[0, 3, 3]+[0, 4, 1]+[0, 4, 2]+[0, 4, 3], \\
{}[0, 4, 3] & = & [0, 0, 4]+[0, 0, 5]+[0, 0, 6]+[0, 0, 7]+[0, 1, 4]+[0, 1, 5]+[0, 2, 3]
\nn & &
+[0, 2, 4]+[0, 2, 5]+[0, 3, 3]+[0, 3, 4]+[0, 4, 2]+[0, 4, 3]+[0, 4, 4], \\
{}[0, 4, 4] & = & [0, 0, 5]+[0, 0, 6]+[0, 0, 7]+[0, 0, 8]+[0, 1, 5]+[0, 1, 6]+[0, 2, 4]
\nn & &
+[0, 2, 5]+[0, 2, 6]+[0, 3, 4]+[0, 3, 5]+[0, 4, 3]+[0, 4, 4]+[0, 4, 5], \\
{}[0,4,l] & = & [0,0,l+1]+[0,0,l+2]+[0,0,l+3]+[0,0,l+4]+[0,1,l+1]
\nn & &
+[0,1,l+2]+[0,2,l]+[0,2,l+1]+[0,2,l+2]+[0,3,l]
\nn & &
+[0,3,l+1]+[0,4,l-1]+[0,4,l]+[0,4,l+1], \label{a040-5}
\eea
where $l\geq 5$.

From \eqref{a000-1}-\eqref{a000-5}, we see that $a_{0,0,l+1}$ is determined by $a_{0,0,l'}$ with $l'\leq l$,
$a_{0,1,l'}$ with $l'\leq l-1$, $a_{0,2,l'}$ with $l'\leq l-1$, $a_{0,3,l'}$ with $l'\leq l-2$, 
and $a_{0,4,l'}$ with $l'\leq l-2$.

From \eqref{a010-1}-\eqref{a010-5}, we see that $a_{0,1,l+1}$ is determined by $a_{0,0,l'}$ with $l'\leq l+1$,
$a_{0,1,l'}$ with $l'\leq l$, $a_{0,2,l'}$ with $l'\leq l-1$, $a_{0,3,l'}$ with $l'\leq l-1$, 
and $a_{0,4,l'}$ with $l'\leq l-2$.

From \eqref{a020-1}-\eqref{a020-5}, we see that $a_{0,2,l+1}$ is determined by $a_{0,0,l'}$ with $l'\leq l+2$,
$a_{0,1,l'}$ with $l'\leq l+1$, $a_{0,2,l'}$ with $l'\leq l$, $a_{0,3,l'}$ with $l'\leq l-1$, 
and $a_{0,4,l'}$ with $l'\leq l-1$.

From \eqref{a030-1}-\eqref{a030-5}, we see that $a_{0,3,l+1}$ is determined by $a_{0,0,l'}$ with $l'\leq l+2$,
$a_{0,1,l'}$ with $l'\leq l+2$, $a_{0,2,l'}$ with $l'\leq l+1$, $a_{0,3,l'}$ with $l'\leq l$, 
and $a_{0,4,l'}$ with $l'\leq l-1$.

We find it more useful to use the following equations obtained by eliminating $[0,0,l+4]$ in
\eqref{a040-1}-\eqref{a040-5} by using \eqref{a000-1}-\eqref{a000-5},
than to use \eqref{a040-1}-\eqref{a040-5}:
\bea
[0,4,0] & = & [0,0,2]+[0,0,3]+[0,2,1]+[0,2,2]+[0,4,0]+[0,4,1],
\\
{}[0,4,1] & = & [0, 0, 3]+[0, 0, 4]+[0, 1, 3]+[0, 2, 2]+[0, 2, 3]
\nn & &
+[0, 4, 0]+[0, 4, 1]+[0, 4, 2],
\\
{}[0, 4, 2] & = & [0, 0, 3]+[0, 0, 4]+[0, 0, 5]+[0, 1, 3]+[0, 1, 4]
\nn & &
+[0, 2, 2]+[0, 2, 3]+[0, 2, 4]+[0, 3, 3]
\nn & &
+[0, 4, 1]+[0, 4, 2]+[0, 4, 3],
\\
{}[0, 4, 3] & = & [0, 0, 4]+[0, 0, 5]+[0, 0, 6]+[0, 1, 4]+[0, 1, 5]
\nn & &
+[0, 2, 3]+[0, 2, 4]+[0, 2, 5]+[0, 3, 3]+[0, 3, 4]
\nn & &
+[0, 4, 2]+[0, 4, 3]+[0, 4, 4],
\\
{}[0, 4, 4] & = & [0, 0, 5]+[0, 0, 6]+[0, 0, 7]+[0, 1, 5]+[0, 1, 6]
\nn & &
+[0, 2, 4]+[0, 2, 5]+[0, 2, 6]+[0, 3, 4]+[0, 3, 5]
\nn & &
+[0, 4, 3]+[0, 4, 4]+[0, 4, 5],
\\
{}[0,4,l] & = & [0,0,l+1]+[0,0,l+2]+[0,0,l+3]+[0,1,l+1]+[0,1,l+2]
\nn & &
+[0,2,l]+[0,2,l+1]+[0,2,l+2]+[0,3,l+3]+[0,3,l+1]
\nn & &
+[0,4,l-1]+[0,4,l]+[0,4,l+1].
\eea
From these, we see that $a_{0,4,l+1}$ is determined by $a_{0,0,l'}$ with $l'\leq l+3$,
$a_{0,1,l'}$ with $l'\leq l+2$, $a_{0,2,l'}$ with $l'\leq l+2$, $a_{0,3,l'}$ with $l'\leq l+1$, 
and $a_{0,4,l'}$ with $l'\leq l$.

In summary, $a_{0,0,l+1}$, $a_{0,1,l+1}$, $a_{0,2,l+1}$, $a_{0,3,l+1}$ and $a_{0,4,l+1}$ are determined
by $a_{0,0,l'}$, $a_{0,1,l'}$, $a_{0,2,l'}$, $a_{0,3,l'}$ and $a_{0,4,l'}$ with $\l'\leq l$.
(These can be determined in the following order: 
$a_{0,0,l+1}$, $a_{0,0,l+2}$, $a_{0,1,l+1}$, $a_{0,1,l+2}$, $a_{0,2,l+1}$, $a_{0,0,l+3}$, 
$a_{0,3,l+1}$, $a_{0,2,l+2}$, $a_{0,4,l+1}$.)
Therefore all the $a_{n,m,l}$ are deteremined in terms of $a_{0,0,0}$, $a_{0,1,0}$, $a_{0,2,0}$, $a_{0,3,0}$,
and $a_{0,4,0}$.

Finally we determine $a_{0,1,0}$, $a_{0,2,0}$, $a_{0,3,0}$, and $a_{0,4,0}$ in terms of $a_{0,0,0}$.
Explicit forms of \eqref{a000-1}, \eqref{a000-2}, \eqref{a000-3} and \eqref{a030-1} are 
\bea
p_4a_{0,0,0} & = & \frac{29}{68}p_2^2a_{0,0,0} + \frac{5040}{17}a_{0,0,1},
\\
p_4a_{0,0,1} & = & \frac{1}{101745}p_2^4a_{0,0,0} + \frac{3159}{7820}p_2^2a_{0,0,1}
   + 1496a_{0,0,2} - \frac{4080}{437}p_2a_{0,2,0},
\\
p_4a_{0,0,2} & = & \frac{1}{450225}p_2^4a_{0,0,1} + \frac{24349}{61380}p_2^2a_{0,0,2}
  + \frac{45486360}{9889}a_{0,0,3}
\nn & &
 + \frac{8}{62031}p_2^3 a_{0,2,0} - 
  \frac{9360}{899}p_2a_{0,2,1} - \frac{13909320}{9889}a_{0,4,0},
\\
p_4a_{0,3,0} & = & -\frac{4}{89001}p_2^3a_{0,1,1} + \frac{36400}{37851}p_2a_{0,1,2}
  + \frac{619}{1276}p_2^2a_{0,3,0} + \frac{81900}{37}a_{0,3,1}.
\eea
Solving these and \eqref{eeq1}-\eqref{eeq7}, we obtain
\bea
a_{0, 1, 0} & = & -\frac{i}{135}p_3a_{0,0,0}, \\
a_{0, 2, 0} & = & \frac{-243 p_2^3 - 1748 p_3^2 + 684 p_2 p_4}{97977600}a_{0,0,0}, \\
a_{0, 3, 0} & = & \frac{i p_3 (207 p_2^3 + 580 p_3^2 - 540 p_2 p_4)}{29099347200}a_{0,0,0}, \\
a_{0, 4, 0} & = & [(37503 p_2^6 + 2834208 p_2^3 p_3^2 + 4492480 p_3^4 - 
      43092 p_2^4 p_4 - 7088256 p_2 p_3^2 p_4
\nn & &
 - 688176 p_2^2 p_4^2 + 
      1373760 p_4^3)/366065131880448000]a_{0,0,0}.
\eea
This shows that $a_{n,m,l}$ are determined by $a_{0,0,0}$, which gives a proof of the conjecture
given in the previous section for $N=4$ case.

\section{Discussion}

We have computed the thermal partition function of SU($N$) $D=2$ matrix quantum mechanics, and from it we 
have obtained some insight into the structure of the spectrum.
We have conjectured that CHS solution is the unique eigenstate of $\text{tr}\;[\Pi^2]$, $\text{tr}\;[\Pi^3], \dots$,
and $\text{tr}\;[\Pi^N]$ if we fix the eigenvalues, and have given proofs of it for $N=3$ and $N=4$ cases.

Our proofs of the uniqueness is too inefficient to apply to higher $N$ cases, and
it is desirable to invent more efficient method for investigating those cases.
From \eqref{mp2}-\eqref{mp4}, we can surmise that the following equation holds for any $n$:
\beq
\text{tr}\,[\Pi^n]\frac{1}{\cal M}f(X)\Big|_{X=Y} = \frac{(-i)^n}{\cal M}
\sum_{i=1}^N(\p_i)^nf(X)\Big|_{X=Y},
\eeq
and if this can be shown, CHS solution is proven to be an eigenfunction of
$\text{tr}\;[\Pi^2]$, $\text{tr}\;[\Pi^3], \dots$, and $\text{tr}\;[\Pi^N]$.

This system has been studied in the view of cut Fock space approach in \cite{k0910,cut}, 
which is a numerically tractable method, and some energy eigenfunctions
have been constructed. Relation between it and our analysis is not immediately clear to us. 
It is important to make the relation clear and develop methods for numerical calculation more.

Although one may think that this system is too simplified for a model of Matrix theory,
it may give hint to the structure of asymptotic plane waves of Matrix theory, which is
important to consider scattering process\cite{petal}.

\renewcommand{\theequation}{\Alph{section}.\arabic{equation}}
\appendix
\addcontentsline{toc}{section}{Appendix}
\vs{.5cm}
\noindent
{\Large\bf Appendix}
\section{SU($N$) group and characters}
\label{appa}
\setcounter{equation}{0}

A Cartan subalgebra $\{H_m;\;m=1,2,\dots,N-1\}$
of Lie algebra {\it su}($N$) can be given as a set of diagonal traceless $N\times N$ matrices.
For example,
\beq
H_m=\frac{1}{\sqrt{m(m+1)}}\left(\begin{array}{@{\,}ccccc@{\,}}
\begin{array}{ccc}
1 & & \\
& \ddots & \\
& & 1 \\
\end{array} & \Bigg\}~m & & & \\
 & -m & & & \\
 & & 0 & & \\
 & & & \ddots & \\
 & & & & 0
\end{array}\right).
\eeq
These satisfy $\text{tr}(H_mH_n)=\delta_{mn}$. Then
a Cartan-Weyl basis is given by
\beq
\{H_m,E_{ij};\;m=1,2,\dots,N-1,\; i,j=1,2,\dots,N,\;i\not\neq j\},
\eeq
where $E_{ij}$ is the matrix whose only nonzero component is at $i$-th row and $j$-th column: $(E_{ij})^i{}_j=1$.
Commutation relations of these operators are
\beq
{}[H_m,H_n]=0,\quad [H_m,E_{ij}]=((\nu^i)_m-(\nu^j)_m)E_{ij},\quad
[E_{ij},E_{kl}]=\delta_{jk}E_{il}-\delta_{il}E_{jk},
\label{suncom}
\eeq
where
\beq
\nu^i=((H_1)^i{}_i,\; (H_2)^i{}_i,\;\dots,\;(H_{N-1})^i{}_i) \quad (i=1,2,\dots, N),
\eeq
are weights of fundamental representation. Note that $\sum_{i=1}^N\nu^i=0$,
and the right hand side of the last equation in \eqref{suncom} can contain $H_m$, because $E_{ii}$ is 
diagonal and can be rewritten in terms of $H_m$.

An ordinary hermitian basis $\{T^a;\;a=1,2,\dots,N^2-1\}$ satisfying $(T^a)^\dagger=T^a$ and 
$\text{tr}(T^aT^b)=\delta^{ab}$
is given by
\beq
\{T^a;\;a=1,2,\dots,N^2-1\}=\{H_m,E_{ij}^+,E_{ij}^-;\;
m=1,2,\dots,N-1,\; 1\leq i<j\leq N\},
\eeq
where
\beq
E_{ij}^+=\frac{1}{\sqrt{2}}(E_{ij}+E_{ji}),\quad E_{ij}^-=\frac{i}{\sqrt{2}}(E_{ij}-E_{ji}).
\eeq
Due to the following:
\beq
\sum_{i,j}(E_{ij})^k{}_l(E_{ji})^{k'}{}_{l'}=\delta^k{}_{l'}\delta^{k'}{}_l,\quad
\sum_{m=1}^{N-1}(H_m)^i{}_j(H_m)^{i'}{}_{j'}=\delta^i{}_j\delta^{i'}{}_{j'}\left(\delta^{ii'}-\frac{1}{N}\right),
\eeq
$T^a$ satisfy 
$\sum_a(T^a)^i{}_j(T^a)^{i'}{}_{j'}=\delta^i{}_{j'}\delta^{i'}{}_j-\frac{1}{N}\delta^i{}_j\delta^{i'}{}_{j'}$,
and therefore
\bea
\sum_a \text{tr}\;(T^aAT^aB) & = & \text{tr}\;(A)\text{tr}\;(B)-\frac{1}{N}\text{tr}\;(AB),
\label{tr1} \\
\sum_a \text{tr}\;(T^aA)\text{tr}\;(T^aB) & = & \text{tr}\;(AB)-\frac{1}{N}\text{tr}\;(A)\text{tr}\;(B).
\label{tr2}
\eea

An element $\theta^aT^a$ of Lie algebra {\it su}($N$) can be expanded in various ways. We define
$\theta^m$, $\theta^{(\pm ij)}$ and $\theta^{(ij)}$ as follows:
\bea
\theta^aT^a & = & \sum_{m=1}^{N-1}\theta^mH_m+\sum_{i<j}(\theta^{(+ij)}E_{ij}^++\theta^{(-ij)}E_{ij}^-) \\
 & = & \sum_{m=1}^{N-1}\theta^mH_m+\sum_{i\not\neq j}\theta^{(ij)}E_{ij}.
\eea

Adjoint representation is given by the tensor product of fundamental and antifundamental representation
with the trace part subtracted. Its representation matrices are given by the structure constant $f_{abc}$:
for a group element $e^{i\lambda^aG^a}$, where $G^a$ are generators and $\lambda^a$ are parameters,
{\it su}($N$) representation $\text{ad}(\lambda)$ and SU($N$) representation $\text{Ad}(\lambda)$
are given by 
\beq
\text{Ad}(\lambda)=e^{i\text{ad}(\lambda)},\quad \text{ad}(\lambda)_{ab}=if_{acb}\lambda^c.
\eeq
When $\lambda^{(ij)}=0$,
$\text{Ad}(\lambda)$ is diagonalized by decomposing the index $a$ into $m$ and $(ij)$:
\beq
\text{Ad}(\lambda)_{mn}=\delta_{mn},\quad \text{Ad}(\lambda)_{(ij)(ji)}=z_j/z_i,
\eeq
where
\beq
z_i=e^{2\pi iy_i},\quad 2\pi y_i=\nu^i\cdot \lambda=\sum_{i=1}^{N-1}(\nu^i)_m\lambda_m.
\label{defzy}
\eeq
Note that $z_i$ or $y_i$ are not independent of each other: $\prod_{i=1}^Nz_i=1$ or 
$\sum_{i=1}^Ny_i=0$.

The character $\chi(R)$ of a representation $R$ is given by a trace of a group element
$g=e^{i\lambda^aG^a}$ over states in $R$:
$\chi(R)=\text{tr}_R(g)$, and we define $\chi(R^k)$ by $\chi(R^k)=\text{tr}_R(g^k)$.
Since $\text{tr}_R(hgh^{-1})=\text{tr}_R(g)$, we can set $\lambda^{(ij)}=0$ without loss of generality.
The characters of $n$-rank symmetric tensor product $\text{Sym}_n(R)$, and
$n$-rank antisymmetric tensor product $\text{Alt}_n(R)$, of a representation $R$ can be
computed by the following Frobenius formulae:
\bea
\chi(\text{Sym}_n(R)) & = & \sum_{\stackrel{\mbox{$\scriptstyle \sum_{k=1}^nki_k=n$}}
{\mbox{$\scriptstyle i_k: \text{nonnegative integer}$}}
}
\prod_{k=1}^n\frac{[\chi(R^k)]^{i_k}}{i_k!\cdot k^{i_k}},\\
\chi(\text{Alt}_n(R)) & = & \sum_{\stackrel{\mbox{$\scriptstyle \sum_{k=1}^nki_k=n$}}
{\mbox{$\scriptstyle i_k: \text{nonnegative integer}$}}
}
(-1)^{n+\sum_{k=1}^ni_k}\prod_{k=1}^n\frac{[\chi(R^k)]^{i_k}}{i_k!\cdot k^{i_k}}.
\eea
The second formula gives zero when $n$ is larger than the dimension of $R$.
The generating functions of these characters is expressed in the following form.
\bea
\sum_{n=0}^{\infty}a^n\chi(\text{Sym}_n(R)) & = & \exp\left(\sum_{n=1}^\infty\frac{a^n}{n}\chi(R^n)\right),\\
\sum_{n=0}^{\dim R}a^n\chi(\text{Alt}_n(R)) & = & \exp\left(-\sum_{n=1}^\infty\frac{(-a)^n}{n}\chi(R^n)\right).
\eea
Often $\chi(R^n)$ is given in the form of $\chi(R^n)=\pm\sum_ib_i^n$, and in this case,
\bea
\sum_{n=0}^{\infty}a^n\chi(\text{Sym}_n(R)) & = & 
\prod_i\exp\left(\pm\sum_{n=1}^\infty\frac{(ab_i)^n}{n}\right)
=\prod_i\exp\left(\mp\log(1-ab_i)\right) \nn
 & = & 
\prod_i(1-ab_i)^{\mp 1},\\
\sum_{n=0}^{\dim R}a^n\chi(\text{Alt}_n(R)) & = & 
\prod_i\exp\left(\mp\sum_{n=1}^\infty\frac{(-ab_i)^n}{n}\right)
=\prod_i\exp\left(\pm\log(1+ab_i)\right) \nn
 & = & 
\prod_i(1+ab_i)^{\pm 1}.
\label{genfunc}
\eea
In the case of SU($N$) adjoint representation, the character is given by 
\beq
\chi(\text{adj.})=\text{tr}(\text{Ad}(\lambda))=\sum_{i,j=1}^N\frac{z_i}{z_j}-1,
\eeq
and 
\bea
\sum_{n=0}^{\infty}a^n\chi(\text{Sym}_n(\text{adj.})) & = & 
(1-a)^{-(N-1)}\prod_{\stackrel{\mbox{$\scriptstyle i,j=1$}}{\mbox{$\scriptstyle i\not\neq j$}}}^N(1-az_i/z_j)^{-1},
\label{chsymadj}\\
\sum_{n=0}^{\dim R}a^n\chi(\text{Alt}_n(\text{adj.})) & = & 
(1+a)^{N-1}\prod_{\stackrel{\mbox{$\scriptstyle i,j=1$}}{\mbox{$\scriptstyle i\not\neq j$}}}^N(1+az_i/z_j).
\label{chaltadj}
\eea

The square of Vandermonde determinant $\prod_{i\not\neq j}(z_i-z_j)$
is the measure for the following orthogonality relation for characters:
\bea
\int d\text{SU}(N)\left(\chi(R')\right)^*\chi(R) & = & \int d\lambda\left(\chi(R')\right)^*\chi(R)
\nn & \equiv &
\frac{1}{N!}\left(\prod_{i=1}^{N-1}\int_0^1dy_i\right)\Big[
\prod_{\stackrel{\mbox{$\scriptstyle i,j=1$}}{\mbox{$\scriptstyle i\not\neq j$}}}^N(z_i-z_j)\Big]
\left(\chi(R')\right)^*\chi(R)
\nn & = &
\delta_{R',R}.
\eea

\newcommand{\J}[4]{{\sl #1} {\bf #2} (#3) #4}
\newcommand{\andJ}[3]{{\bf #1} (#2) #3}
\newcommand{\AP}{Ann.\ Phys.\ (N.Y.)}
\newcommand{\MPL}{Mod.\ Phys.\ Lett.}
\newcommand{\NP}{Nucl.\ Phys.}
\newcommand{\PL}{Phys.\ Lett.}
\newcommand{\PR}{Phys.\ Rev.}
\newcommand{\PRL}{Phys.\ Rev.\ Lett.}
\newcommand{\PTP}{Prog.\ Theor.\ Phys.}
\newcommand{\hepth}[1]{{\tt hep-th/#1}}
\newcommand{\arxivhep}[1]{{\tt arXiv.org:#1 [hep-th]}}

\end{document}